\documentclass{IEEEtran}
\usepackage{cite}
\usepackage{amsmath,amssymb,amsfonts}
\usepackage{graphicx}
\usepackage{textcomp}
\usepackage{algorithm}  
\usepackage{algpseudocode}  
\usepackage{amsmath}  
\usepackage{textcomp}
\usepackage{multirow}
\newcommand{\degree}{^\circ}
\usepackage{color}
\usepackage{arydshln}
\usepackage{nicematrix}
\usepackage{tikz}
\usetikzlibrary{matrix,fit}

\def\BibTeX{{\rm B\kern-.05em{\sc i\kern-.025em b}\kern-.08em
    T\kern-.1667em\lower.7ex\hbox{E}\kern-.125emX}}
\begin{document}
\title{{\fontsize{24}{26}\selectfont{Communication\rule{29.9pc}{0.5pt}}}\break\fontsize{16}{18}\selectfont
A Parallel Block Preconditioner Based VIE-FFT Algorithm for Modeling Electromagnetic Response from Nanostructures}
\author{Chengnian Huang and Wei E.I. Sha
\thanks{ This work was supported by the National Natural Science Foundation of China under Grant 61975177 and Grant U20A20164. (\emph{Corresponding author: Wei E. I. Sha.})}
\thanks{Chengnian Huang and Wei E. I. Sha are with the College of Information Science and Electronic Engineering, Zhejiang University, Hangzhou 310027, China. (e-mail: 12031047@zju.edu.cn; weisha@zju.edu.cn).}}
\maketitle
\begin{abstract}
The superior ability of nanostructures to manipulate light has propelled extensive applications in nano-electromagnetic components and devices. Computational electromagnetics plays a critical role in characterizing and optimizing the nanostructures. In this work, a parallel block preconditioner based volume integral equation (VIE)-fast Fourier transform (FFT) algorithm is proposed to model the electromagnetic response from representative nanostructures. The VIE using uniform Cartesian grids is first built, and then the entire volumetric domain is partitioned into geometric subdomains based on the regularity and topology of the nanostructure. The block diagonal matrix is thus established, whose inverse matrix serves as a preconditioner for the original matrix equation. The resulting linear system is solved by the bi-conjugate gradient stabilized (BiCGSTAB) method with different residual error tolerances in the inner and outer iteration processes; and the FFT algorithm is used to accelerate the matrix-vector product (MVM) operations throughout. Furthermore, because of the independence between the inner processes of solving block matrix equations, the OpenMP framework is empolyed to execute the parallel operations. Numerical experiments indicate that the proposed method is effective and reduces both the iteration number and the computational time significantly for the representative nano-electromagnetic problems like the dielectric focusing metasurfaces and the plasmonic solar cells.
\end{abstract}


\section{Introduction}
Nanostructures serve as the fundamental building blocks for the nano-electromagnetic designs and hold significant importance in the emerging engineering applications, such as virtual/augmented reality\cite{1VR2021}, metalens\cite{2metalenses2017}, organic solar cells\cite{3osc2005} and photonic integrated circuits\cite{4circuits2017}, etc. Computational electromagnetics plays a vital role in characterizing and optimizing nanostructures, effectively reducing the costs and time associated with realistic experimental fabrication. In particular, the accurate modeling of light-matter interaction within nanostructures is essential. And this modeling endeavor aims to propel the exploration of novel physical effects and their corresponding experimental investigation. Numerous research studies have been conducted to address categories of nanostructure problems, such as the planar plasmonic structures\cite{5planar2022}, complex disordered stacks of gold nanorods or three-dimensional photonic crystals\cite{6Uplas2014}, nonlinear optical process of metal nanoparticles\cite{7Nonlinear2008}, etc. In this work, the focusing metasurfaces and the plasmonic solar cells are chosen as representative examples, which possess commonly used structural features that display far-field interference and near-field coupling effects, respectively.

Considering the dimensions and material composition of nanostructures, there are three kinds of popular rigorous methods including the differential-equation-based (DE) methods, the integral-equation-based (IE) methods, and the semi-analytical methods. The semi-analytical methods can only solve the electromagnetic response for specific nanostructures. The DE methods involving the finite-difference time-domain (FDTD) method\cite{8fdtd}, the finite-difference frequency-domain (FDFD) method\cite{9veronis2007fdfd} and the finite element method (FEM)\cite{10jin2015fem} discretize the whole region, resulting in a large number of unknowns to be solved. Moreover, the accuracy of modeling high-contrast plasmonic structures with strong evanescent wave coupling is reduced due to the dispersion error in both the FEM and the FDTD method, as well as the staircase approximation used in the FDTD method\cite{11hes2012}. Differently, the IE methods only need to discretize the object. Not only is the number of unknowns small, but also the radiation boundary condition can be satisfied automatically by using the dyadic Green's function. They usually have higher accuracy. Because nanostructures often exhibit complex and intricate geometries including sharp edges, corners and fine details, etc, some of which involve open surfaces, the IE methods are the superior option compared with the DE methods. The IE methods are classified into the surface integral equation (SIE) method\cite{12kern2009sie} and the volume integral equation (VIE) method\cite{13ewe2007vie}. For the metallic, homogeneous dielectric and composite metallic and dielectric objects, the SIE is preferred to be established at the surface of the nanostructure\cite{14chew2022}, but the precorrected-FFT algorithm\cite{15pre-fft} is required for the near-field calculations. In comparison, the VIE in conjunction with the method of moments (MoM)\cite{16harrington1993mom} is an easy-to-implement and flexible method to calculate the electromagnetic scattering from dielectric bodies of arbitrary shape and  inhomogeneous material composition\cite{17pxm2018}. Although for some scattering structures the impedance matrix can be poorly conditioned and the conventional MoM suffers from tremendously high computational cost and memory requirement, efficient iterative and fast algorithms have alleviated this problem to some extent.

In recent years, many efforts have been made to apply various iterative and fast algorithms to reduce the complexity and memory cost of the MoM solution. The commonly used iterative approach to solve the VIE is the bi-conjugate gradient stabilized (BiCGSTAB) method\cite{18bicgstab} from the Krylov subspace family. The Krylov methods require the computation of some matrix-vector product (MVM) operations at each iteration, which accounts for the major computational cost of this class of methods. However, by performing the MVM with the 3-D fast Fourier transform (FFT)\cite{19FFT}, the computational complexity and the memory use are reduced to $O(N$log$N)$ and $O(N)$, respectively. Normally, the FFT algorithm requires the volume of the object to be discretized into uniform hexahedral cells in order to use the Toeplitz property of the impedance matrix, which results in a large number of unknowns for accurate geometric model with the staircase approximation. However, the near-field calculation without precorrection can save a great amount of computer resources, especially for the nano-electromagnetic problems involving strong near-field interactions that cover a large range. There is another problem that the typical MoM implementations for dielectric bodies do not consider the induced currents flowing between the dielectric volumes and the free space, which has been discussed in \cite{20FFT-MLFMM2021}. But this work primarily focused on capturing the field interaction effect of unit cells, manipulating an efficient preconditioner and conducting further studies on the accuracy and efficiency of the proposed preconditioned solver. More extensions of this problem will be made in our future work. 

In order to accelerate the convergence rate of a Krylov method, substantial efforts have been devoted to the development of the straightforward preconditioners. The commonly-adopted preconditioning tools include the incomplete LU factorization\cite{21ILU} and the symmetric successive over-relaxation (SSOR)\cite{22SSOR}, both of which are difficult to implement in parallel. Therefore, we propose a combination of parallel computing techniques and an efficient block preconditioning method inspired by the rank-revealing decomposition preconditioner\cite{23pissoort2006rank}, which is well-suited for the organized unit nanostructures. Based on the fact that the spectral properties of the impedance matrix are mainly determined by the near-field dependence of the integral-equation kernel especially for nano-electromagnetic applications\cite{24near kernel2004}, the entire volumetric domain is partitioned into some geometric subdomains. As a result, a block diagonal matrix resulting from the local geometries of near-field interactions is established to serve as an approximation of the original impedance matrix. The inversion of the approximate matrix is referred to as a preconditioner, and the process requires solving the block matrix equations simultaneously. Therefore, the BiCGSTAB-FFT method is once again adopted to solve the submatrix equations in the inner iterations, combined with the OpenMP parallel technique for loops in the outer iterations.

In this paper, the theory of the preconditioned VIE-FFT algorithm is described in Section $\mathrm{\uppercase\expandafter{\romannumeral2}}$. In Section $\mathrm{\uppercase\expandafter{\romannumeral3}}$, two commonly encountered regular nanostructures are provided to demonstrate the correctness and efficiency of the proposed method. Finally, the conclusion is given in Section $\mathrm{\uppercase\expandafter{\romannumeral4}}$.

\section{THEORY}
\subsection{VIE-MoM}
Here we consider an inhomogeneous 3-D nanostructure illuminated by a plane wave at a specific frequency of interest.  $\boldsymbol{E}^{s}$ is the corresponding scattered field due to the induced volumetric polarization current $\boldsymbol{J}^{s}$ as follows\cite{25MF1989}:\\
\begin{equation}
	\begin{aligned}
		\boldsymbol{E}^{s}=&-j \omega \mu_{0} \int_{V} g\left(\boldsymbol{r}, \boldsymbol{r}^{\prime}\right) \mathbf{\boldsymbol{J}}^{s}\left(\boldsymbol{r}^{\prime}\right) dv^{\prime}\\
		\;&+\frac{1}{j \omega \varepsilon_{0}} \int_{V} \nabla \nabla g\left(\boldsymbol{r}, \boldsymbol{r}^{\prime}\right) \cdot\boldsymbol{J}^{s}\left(\boldsymbol{r}^{\prime}\right) dv^{\prime}\\
		=&-j \omega \mu_{0} \int_{V} \overleftrightarrow{\mathbf{G}}\left(\boldsymbol{r}, \boldsymbol{r}^{\prime}\right) \cdot\boldsymbol{J}^{s}\left(\boldsymbol{r}^{\prime}\right) d v^{\prime}
	\end{aligned}
\end{equation}
where $g\left(\boldsymbol{r}, \boldsymbol{r}^{\prime}\right)=e^{-jk\left|\boldsymbol{r}-\boldsymbol{r}^{\prime}\right|} / 4 \pi\left|\boldsymbol{r}-\boldsymbol{r}^{\prime}\right|$ denotes the scalar Green’s function in free space, $\overleftrightarrow{\boldsymbol{G}}$ denotes the corresponding dyadic Green's function, $\boldsymbol{r}$ and $\boldsymbol{{r}}^{\prime}$ are the observation and source point locations. The permittivity and permeability in free space are denoted by $\varepsilon_{0}$ and  $\mu_{0}$, respectively.

According to the relation between the polarization current and the electric polarization vector, the total electric field could be described by $\boldsymbol{J}^{s}$ as
\begin{equation}
	\boldsymbol{E}^{\text {tot }}=\frac{\boldsymbol{J}^{s}(r)}{j \omega\left(\varepsilon-\varepsilon_{0}\right)}.
\end{equation}

The total electric field is the summation of the incident field and the scattered field, and thus the volume integral equation can be written as 
\begin{equation}
	\boldsymbol{E}^{t o t}=\boldsymbol{E}^{i n c}+\boldsymbol{E}^{s}.
\end{equation}

Based on the equations above, one can explicitly write the equation with the unknown $\boldsymbol{J}^{s}$ as
	\begin{equation}
		\frac{\boldsymbol{J}^{s}(\boldsymbol{r})}{j \omega\left(\varepsilon-\varepsilon_{0}\right)}+j \omega \mu_{0} \int_{V} \overleftrightarrow{\boldsymbol{G}}\left(\boldsymbol{r}, \boldsymbol{r}^{\prime}\right) \cdot \boldsymbol{J}^{s}\left(\boldsymbol{r}^{\prime}\right) d v^{\prime}=\boldsymbol{E}^{i n c}.
\end{equation}

Here, the MoM is used to discretize the equation. The unknown volumetric currents can be expanded into sets of basis functions, then the problem is converted to minimize the associated residual errors with sets of weighting or testing functions \cite{26chen2007weight}. Considering our examples have regular geometries that allow low-cost and easy fabrication, a hexahedron mesh is utilized to discretize the volumetric structure, enabling the fast MVM and the efficient preconditioning. The unknown currents are expanded into rooftop basis functions
\begin{equation}
	\boldsymbol{J}^{s}=\sum_{i=1}^{3} \boldsymbol{p}_{i} \sum_{k=0}^{N_{1}} \sum_{m=0}^{N_{2}} \sum_{n=0}^{N_{3}} J_{i}^{D}(k m n) T_{k m n}^{s}.
\end{equation}
The  $\boldsymbol{p}_{i}$  are the direction vectors, $J_{i}^{D}$ are the coefficients of current basis functions, and $T_{k m n}^{s}$ are the volumetric rooftop functions. Then, the pulse function is applied to test the equation. Considering the Cartesian coordinate system, all three polarization $x$-, $y$-, and $z$-components of the unknown currents are taken into account. The above equation can be transformed into the matrix equation
	\begin{equation}
			\underbrace{\left(\begin{array}{ccc}
			L_{xx} & L_{xy} & L_{xz}  \\
			L_{yx} & L_{yy} & L_{yz} \\
			L_{zx} & L_{zy} & L_{zz}
		\end{array}\right)}_A\underbrace{\left(\begin{array}{c}
			{J}_{x}^{s}\\
			{J}_{y}^{s}\\
			{J}_{z}^{s}
		\end{array}\right)}_x=\underbrace{\left(\begin{array}{c}
			{E}_{x}^{inc}\\
			{E}_{y}^{inc}\\
			{E}_{z}^{inc}
		\end{array}\right)}_b
	\label{eq6}
	\end{equation}
	where 
	\begin{equation}
		L_{i j}= \begin{cases}L_{i i}^c+L_{i i}^q, & \text { for } i=j \\ L_{i j}^q, & \text { for } i \neq j\end{cases}
	\end{equation}
	$L_{i i}^c{J}_{i}^{s}$ and $L_{i j}^q{J}_{i}^{s}$ are defined as
	\begin{equation}
		L_{i i}^c J_i^s= \frac{J_i^s(\boldsymbol{r})}{j \omega\left(\varepsilon-\varepsilon_{0}\right)}-j \omega \mu_{0}\int_v J_i^s\left(\boldsymbol{r}^{\prime}\right) g\left(\boldsymbol{r}, \boldsymbol{r}^{\prime}\right) d v^{\prime}
	\end{equation}
	\begin{equation}
		L_{i j}^q J_i^s=\frac{1}{j \omega \varepsilon_{0}} \frac{\partial}{\partial u_i} \int_v \frac{\partial J_j^s\left(\boldsymbol{r}^{\prime}\right)}{\partial u_j^{\prime}} g\left(\boldsymbol{r}, \boldsymbol{r}^{\prime}\right) d v^{\prime}.
\end{equation}

\begin{figure}[!t]
	\centerline{\includegraphics[width=\columnwidth]{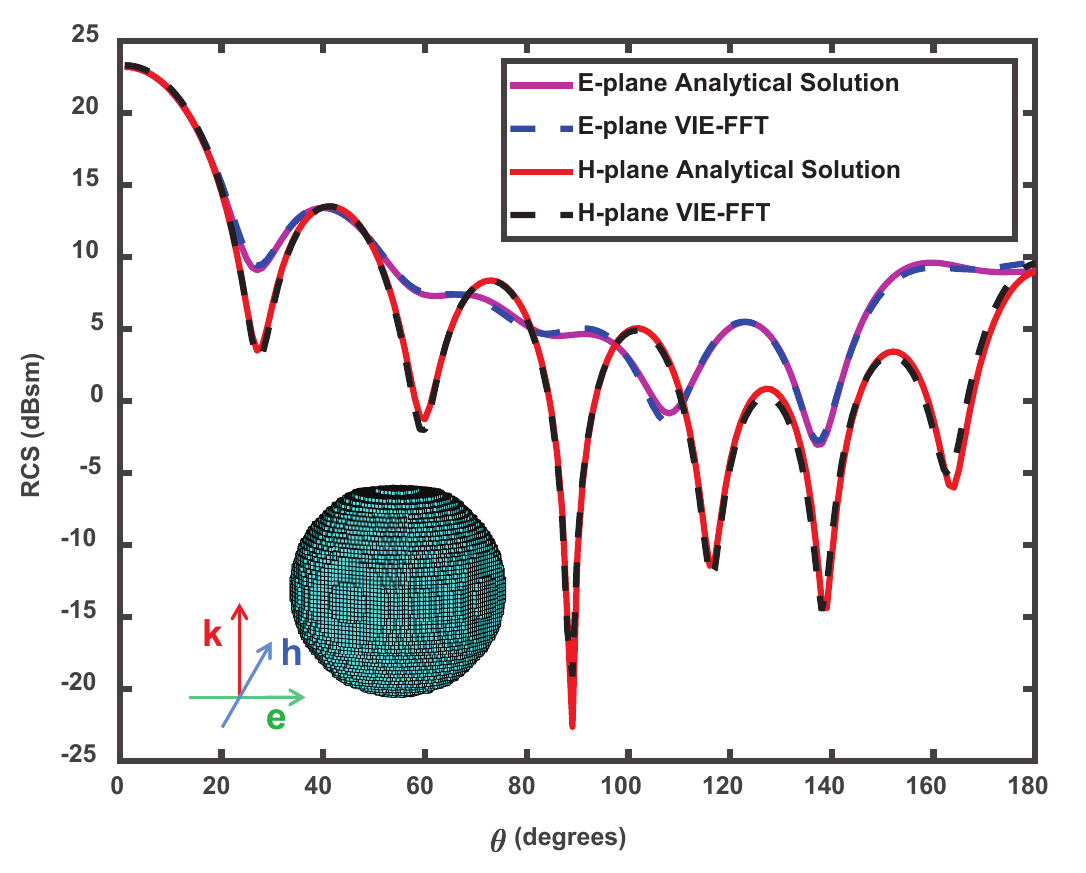}}
	\caption{Bistatic RCS of a dielectric sphere with the radius of $r$ = 400 nm at 750 THz. ($\epsilon_{r}$ = 2.25)}
	\label{huang1}
\end{figure}
\subsection{The Preconditioned Iterative Solver}
Discretizing the VIE with MoM, a dense impedance matrix $\boldsymbol{A}$ will be generated as shown in \eqref{eq6}. Considering the symmetry of the linear system, some algorithms like the GMRES\cite{27gmres}, the CGS\cite{28cgs} and the BiCGSTAB are all appliable, which have low memory cost and good convergence properties as well. Besides, they are easy to combine with a preconditioner. In this work, the BiCGSTAB method is employed as a proof of principle, which shows good performances combined with the parallel block preconditioner. The formulation is equivalent to applying the BiCGSTAB method to the explicitly preconditioned system: $\boldsymbol{\widetilde{A}}\widetilde{\boldsymbol{x}}= \widetilde{\boldsymbol{b}}$, where $\boldsymbol{\widetilde{A}}=\boldsymbol{{K}_{1}^{-1} A K_{2}^{-1}},\,\boldsymbol{\widetilde{x}}=\boldsymbol{K}_{2} \boldsymbol{x}$, and $\boldsymbol{\widetilde{b}}=\boldsymbol{K}_{1}^{-1} \boldsymbol{b}$. We consider a right preconditioner, which means that $\boldsymbol{K}_1$ is the unit matrix, and $\boldsymbol{K}_2$ is an approximation of the impedance matrix $\boldsymbol{A}$. In the outer iteration process, two extra matrix equations $\boldsymbol{y}=\boldsymbol{K}_{2}^{-1}\boldsymbol{p}_{i}$ and $\boldsymbol{z}=\boldsymbol{K}_{2}^{-1}\boldsymbol{s}$ need to be solved, where $\boldsymbol{y}$, $\boldsymbol{p}_{i}$, $\boldsymbol{z}$ and $\boldsymbol{s}$ correspond to the vectors involved in this process. For the purpose of reducing computational complexity, we explore the structure for constructing the block preconditioners, that is to say, $\boldsymbol{K}_{2}\boldsymbol{x}=\boldsymbol{b}$ (representing the two extra equations above) could be approximately viewed as solving the combination of matrix equations of divided blocks, e.g., $\boldsymbol{A}_{1}\boldsymbol{x}=\boldsymbol{b}_{1}$, $\boldsymbol{A}_{2}\boldsymbol{x}=\boldsymbol{b}_{2}$, etc. Since the interaction between blocks is not considered, only the diagonal of the large matrix has nonzero values. The extra matrix equation to be solved is given as follows:
\begin{equation}
\left(\begin{array}{l}
			\boldsymbol{A}_1, 0, \ldots \\
			0, \boldsymbol{A}_2, \ldots \\
			\ldots\ldots\ldots \\
			0, \ldots, \boldsymbol{A}_N
		\end{array}\right)\left(\begin{array}{l}
			\boldsymbol{J}_1 \\
			\boldsymbol{J}_2 \\
			\vdots \\
			\boldsymbol{J}_N
		\end{array}\right)=\left(\begin{array}{l}
			\boldsymbol{b}_1 \\
			\boldsymbol{b}_2 \\
			\vdots \\
			\boldsymbol{b}_N
		\end{array}\right).
	\label{eq10}
\end{equation}
In view of the vector current and the interaction of currents along the three directions, matrices including coefficients of the interactive $x$-, $y$-, and $z$-components are established. Here is an example of the impedance matrix for a divided block:
\begin{equation}
	\boldsymbol{A}_1= \left(\begin{tikzpicture}[baseline=(current bounding box.center)]
		\tikzstyle{every node}=[font=\small]
	\matrix (m) [matrix of math nodes, nodes in empty cells] {
		A_{1x1x},&\ldots,&A_{1xnx},&A_{1x1y},& \ldots,&A_{1x1z},&\ldots,\\
		\vdots &&&&&&\\
		A_{nx1x},&\ldots, &A_{nxnx}, &A_{nx1y}, &\ldots, &A_{nx1z},&\ldots\\
		A_{1y1x},&\ldots, &A_{1ynx}, &A_{1y1y}, &\ldots,&A_{1y1z},&\ldots\\
		\vdots \\
		A_{ny1x},&\ldots, &A_{nynx},&A_{ny1y}, &\ldots,&A_{ny1z},&\ldots\\
		A_{1z1x},&\ldots, &A_{1znx},&A_{1z1y}, &\ldots,&A_{1z1z},&\ldots\\
		\vdots \\
		A_{nz1x},&\ldots, &A_{nznx},&A_{nz1y}, &\ldots,&A_{nz1z},&\ldots\\
	};
	\node[draw,very thick,dashed,fit=(m-1-1)(m-1-3)(m-3-1)(m-3-3)] {};
\end{tikzpicture}\right).
\end{equation}
\begin{figure}[!t]
	\centerline{\includegraphics[width=\columnwidth]{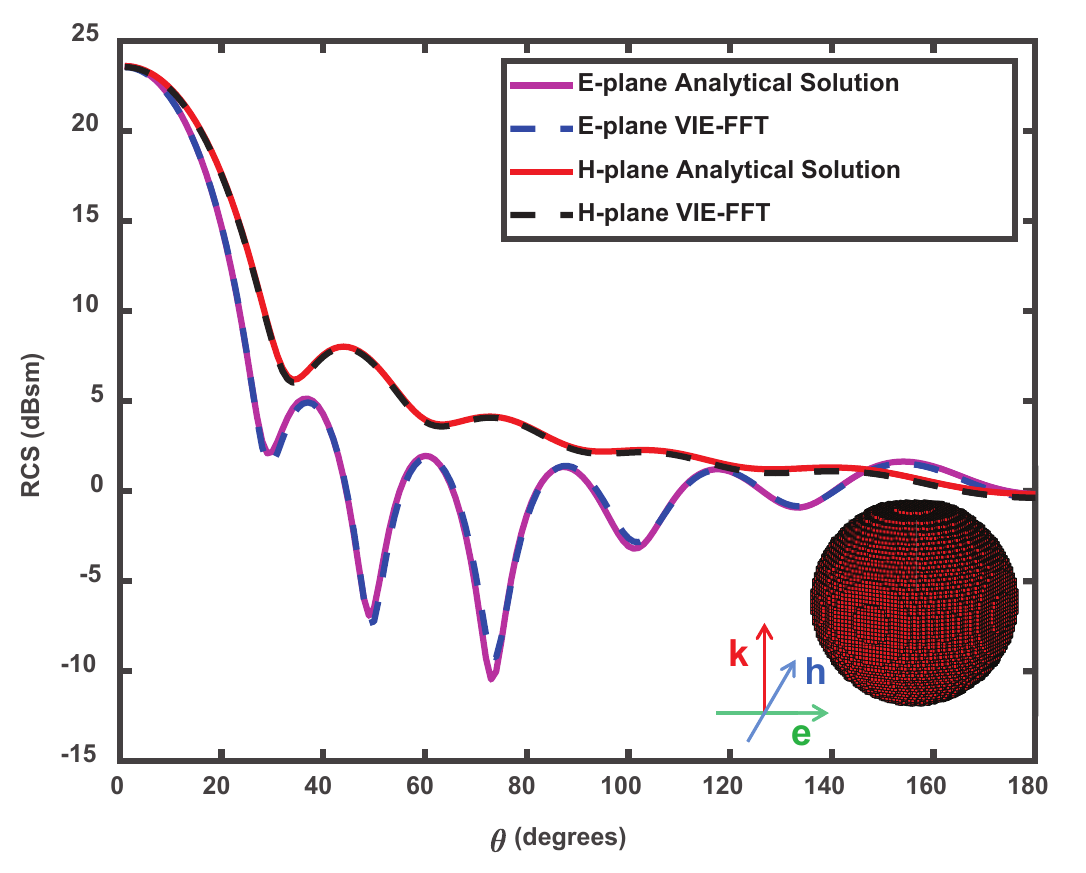}}
	\caption{Bistatic RCS of  a metallic sphere with the radius of $r$ = 400 nm at 750 THz. ($\epsilon_{r}$ = $-$1.05 $-$ $j$5.62)}
	\label{huang2}
\end{figure}
The coefficients of $x$-components in the dotted box correspond to the operator $L_{xx}$ in \eqref{eq6},  which represent the contribution of the currents of $x$-components to the electric fields of $x$-components, likewise for the $y$- and $z$-components. For a divided block, solving the submatrix equation  is an inner iteration process, which also uses the BiCGSTAB-FFT method. Because only the approximate inverse of matrix $\boldsymbol{A}$ needs to be considered, the submatrix equation in \eqref{eq10} can be solved with a larger residual error, i.e., the stopping criteria of iteration can be soften. After retaining the newly solved current vector of the corresponding block, all currents are arranged into a new vector to continue the outer iteration process. Notably, since the interaction between blocks is not considered, data exchange between different blocks is not necessary, which is ideal for parallel computing. The OpenMP parallelization paradigm provides a multithread capacity and fully makes use of the features of shared memory\cite{29xuben2012parallel}, which executes multiple threads for loops of solving block matrix equations. In general, the number of threads is often comparable to the number of divided blocks. It is clear that if there are more blocks to be divided, the time to solve the matrix equation for each block will be shortened. Nevertheless, the use of smaller-sized diagonal block matrices leads to a poorer approximation of the matrix $\boldsymbol{A}$, thus hindering the efficiency of the outer iteration process. Evidently, the division of blocks significantly influence the efficiency of the outer iteration, resulting in a trade-off between the time required for the solution of submatrix equations in one iteration and the number of outer iteration steps. 

\begin{figure}[!t]
	\centerline{\includegraphics[width=\columnwidth]{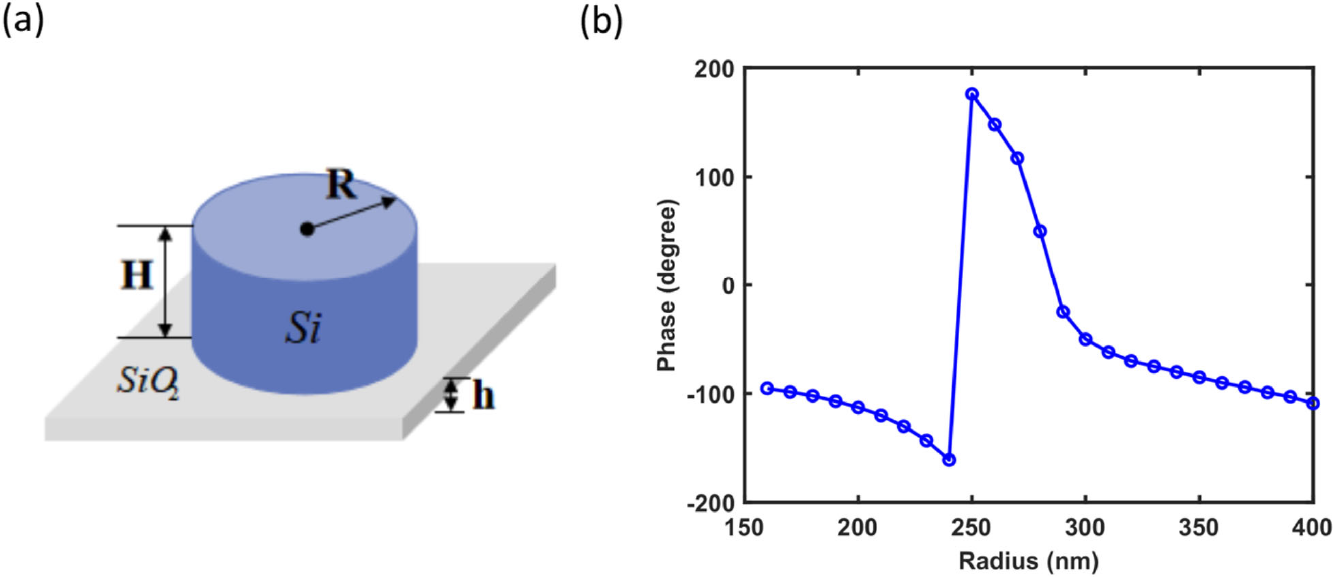}}
	\caption{The change of transmission phase with variation of cylinder’s radius (a) the schematic of single $\mathrm{Si}$ cylinder with substrate of $\mathrm{SiO}_{2}$: $H$ = 250 nm, $h$ = 25 nm. (b) the diagram of phase change.}
	\label{huang3}
\end{figure}
\section{Numerical Results}
In this section, some numerical results are shown to illustrate the effectiveness of the VIE-FFT algorithm with the preconditioner. Firstly, in order to verify the accuracy of the algorithm precisely, we consider the scattering of a dielectric sphere and a metallic sphere with the radius of $r$ = 400 nm (5,832,000 unknowns at 750 THz). The relative permittivity of the dielectric material is 2.25 corresponding to the refractive index of $n=1.5$. The complex permittivity of the metallic material is $\epsilon_{r}$ = $-$1.05 $-$ $j$5.62. The comparison to the Mie series solution\cite{30mie1986} is made for the radar cross section (RCS)  of the sphere as  plotted in Fig. \ref{huang1} and Fig. \ref{huang2}. The relative L2-norm errors of the E-plane RCS are $2.5\%$ and $3.75\%$, respectively. It can be found that there is an excellent agreement between them.

Next, we investigate the performance of the preconditioned algorithm for modeling electromagnetic response from two examples of nanostructures, which are performed with an IBM server of 256 GB memory. 

\subsection{Focusing dielectric metasurfaces}   
\begin{figure}[!t]
	\centerline{\includegraphics[height=0.37\textwidth,width=\columnwidth]{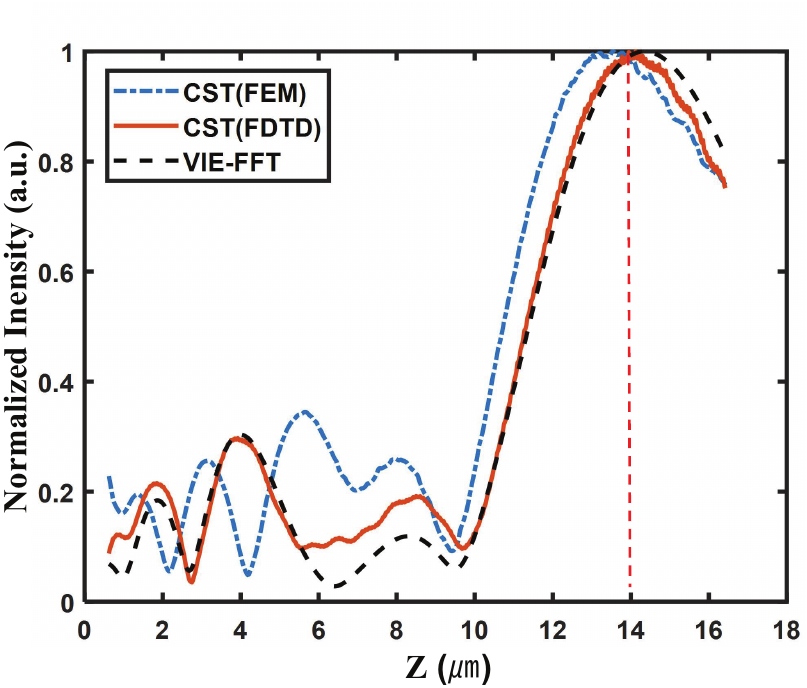}}
	\caption{The normalized electric-field amplitude of the vertical cut of the focusing spot size at $x$ = 0 by the proposed algorithm and CST software. The focal plane of the metasurface is designed to be $d$ = 14 $\mu$m.}
	\label{huang5}
\end{figure}

\begin{figure}[t]
	\centerline{\includegraphics[width=\columnwidth,height=0.368\textwidth]{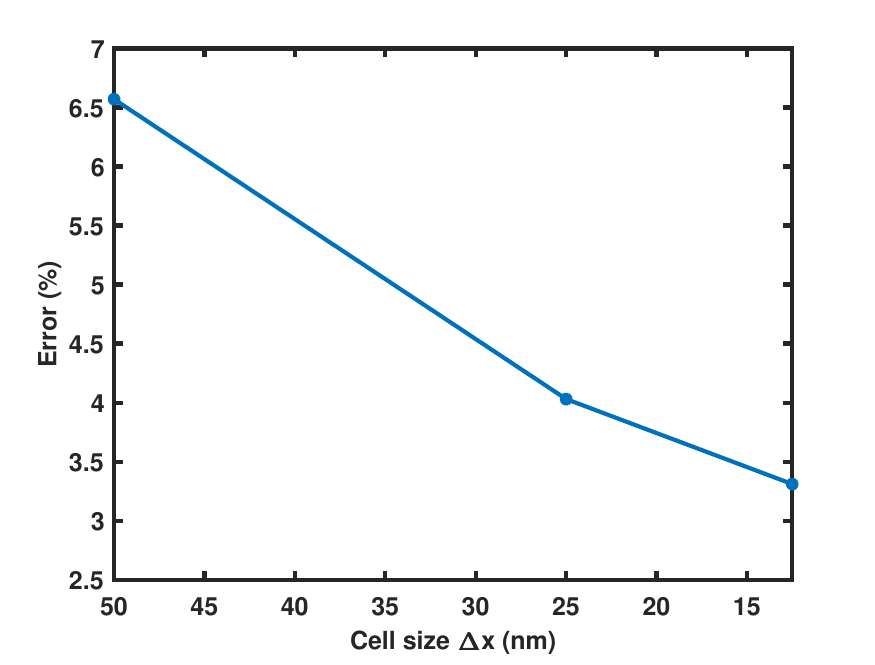}}
	\caption{Convergence of the preconditioned VIE-FFT algorithm: the  relative L2-norm errors with different cell sizes (12.5, 25, and 50 nm). The solution of the FDTD method of CST is taken as the numerical reference.}
	\label{huang6}
\end{figure}

Electromagnetic metasurface is a useful structure to control the beam propagation by phase tailoring. In order to reduce the loss at optical frequencies and increase the feasibility of fabrication, extensive studies have been conducted on the dielectric-only metasurfaces. Here, the resonating disk  as a unit cell is made of silicon with the permittivity of $\epsilon_{r}=12.25$, placed onto a substrate of $\epsilon_{r}=2.25$.  The thicknesses of $\mathrm{Si}$ and $\mathrm{SiO}_{2}$ are 250 nm and 25 nm, respectively. The schematic of the cylinder is displayed in Fig. \ref{huang3}(a). Then, we study the transmission phase of the disk as a function of radius, where the working frequency is fixed at 214 THz. As shown in Fig. \ref{huang3}(b), complete $360\degree$ transmission phase is covered while the radius ranges from 160 nm to 400 nm. According to the phase profile settings: \\
\begin{equation}
	\Phi(x, y)=k_{\mathrm{SiO}_{2}}\left(\sqrt{x^{2}+y^{2}+d^{2}}-d\right)
\end{equation}
\begin{figure}[!t]
	\centerline{\includegraphics[width=\columnwidth]{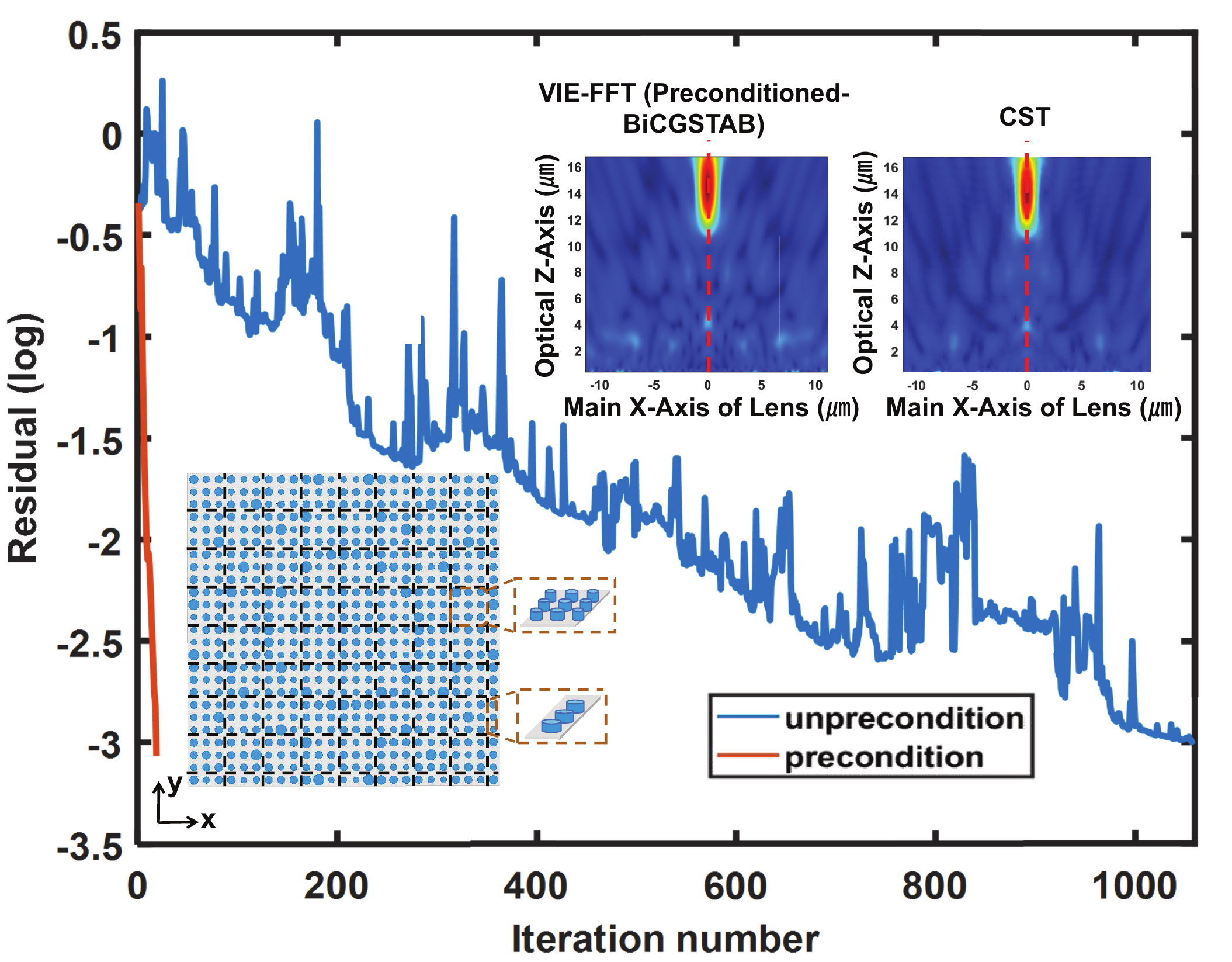}}
	\caption{Residual error versus iteration number for the metasurface. The inset shows the simulated intensity profiles by the preconditioned VIE-FFT algorithm and CST software.}
	\label{huang4}
\end{figure}\\
where $k_{\mathrm{SiO}_{2}}$ is the wave number in the dielectric, $x$, $y$ are the coordinate positions of the units on the plane, $d$ is the focal length, and the size of unit cell is 900 nm. A metasurface for plane-wave focusing could be designed by arranging the disks of different sizes at a fixed focal length. The structure is designed to length $\times$ width $\times$ height as 16$\lambda_0$ $\times$ 16$\lambda_0$ $\times$ 0.2$\lambda_0$ ($\lambda_0$ is the wavelength of free space) and creating 47 grids per $\lambda_0$ (considering high dielectric contrast), incident wave is polarized along the $x$-direction, and the focal length $d$ is set to $10\lambda$, i.e., 14 $\mu$m. So the total number of mesh elements is 6,187,500 and that of unknown currents along the three dimensions is 18,562,500. Nearly 6 GB memory cost is required. To employ the parallel block preconditioner, we divide the structure evenly into blocks based on its regularity of  arrangements. The residual error of the outer iteration is set to 0.001, whereas the inner residual error for the stopping criteria of preconditioning is 0.01. In the Fig. \ref{huang4}, the distribution of the electric-field amplitude in the $x$-$z$ propagation plane is shown, along with the results simulated by the commercial software CST. The performance of convergence is also displayed. A relatively small number of iterations are needed for the proposed method, whereas the original un-preconditioned VIE-FFT algorithm fails to converge to $\epsilon=10^{-3}$ within 1000 iterations. To validate the accuracy of results quantitatively, the normalized electric-field amplitude of the vertical cut of the focusing spot size at $x$ = 0 is compared in Fig. \ref{huang5}. We can see that the field results from the VIE method and the FDTD method of CST are in good agreement. They produce the accurate focusing position at 14 $\mu$m as the theoretical design, whereas the focusing position calculated by the FEM of CST slightly shifts. In addition, the electric-field amplitude calculated by CST shows instability near the focusing position, which might contribute to the error to some extent. Therefore, the convergence property is studied to re-examine the reliability of the proposed method. The structure is discretized  with different cell sizes (12.5, 25, and 50 nm, $\Delta$$x$ = $\Delta$$y$ = $\Delta$$z$). As the grid size decreases from 50 to 12.5 nm, equivalently, the number of grids increases from 1,215,000 to 71,280,000. We take the results simulated by the FDTD method  of CST with a finer grid as the reference, then compute corresponding relative L2-norm errors as depicted in Fig. \ref{huang6}. It can be observed that the error becomes smaller as the grid size decreases, which demonstrates the stability and accuracy of the method. Evidently, sufficient mesh cells are required for good convergence. Furthermore, Table \ref{table1} shows the total calculation time of different methods. 
\begin{table}
	\scriptsize 
	\begin{center}		
		\caption{COMPUTATIONAL STATISTICS OF METASURFACE WITH THE RESIDUAL ERROR 
		\label{table1}
		$\epsilon=10^{-3}$}
		\setlength{\tabcolsep}{0.8mm}
		\begin{tabular}{|c|c|c|c|c|}
			\hline \multirow{4}{*}{  }& \multicolumn{4}{|c|}{ CPU time} \\
			\cline { 2 - 5 } Number & CST& CST& VIE & preconditioned VIE\\
			\newline  { of units }  & (FEM) &  (FDTD)  & method & method \\
			\hline $25 \times 25$ & 9 h 46 min 47 s  & 5 h 6 min 10 s & 10 h 28 min 39 s & 3 h 35 min 13 s\\
			\hline
		\end{tabular}
	\end{center}
\end{table}
We can see that compared to the FDTD method, the FEM and the unpreconditioned VIE method are slower. After introducing the preconditioner, the overall CPU time is roughly reduced to one third of the unpreconditioned one. Compared with the CST, the proposed preconditioned method shows a substantial speedup. We have also tried alternative ways of divisions and find that there is no significant difference for the CPU time except for the slight speed improvement brought by the finer division of blocks, which would be attributed to the electric and magnetic resonance modes dominated by the strong dipole moment located in the center of the nanopillar\cite{31dipole2015}. The field information contained in each block is rich enough to represent the characterization of the whole matrix.
\begin{figure}
	\centerline{\includegraphics[width=\columnwidth]{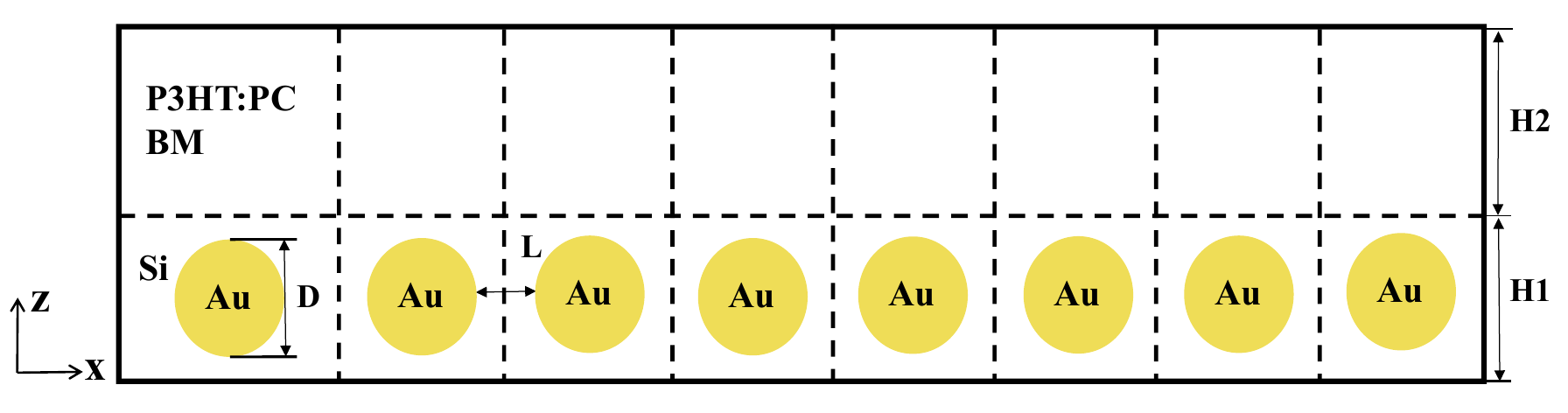}}
	\caption{The schematic diagram of plasmonic solar cells with uniform blocks.}
	\label{huang7}
\end{figure}
\subsection{Plasmonic solar cells}   
\begin{figure}[t!]
	\centerline{\includegraphics[width=\columnwidth]{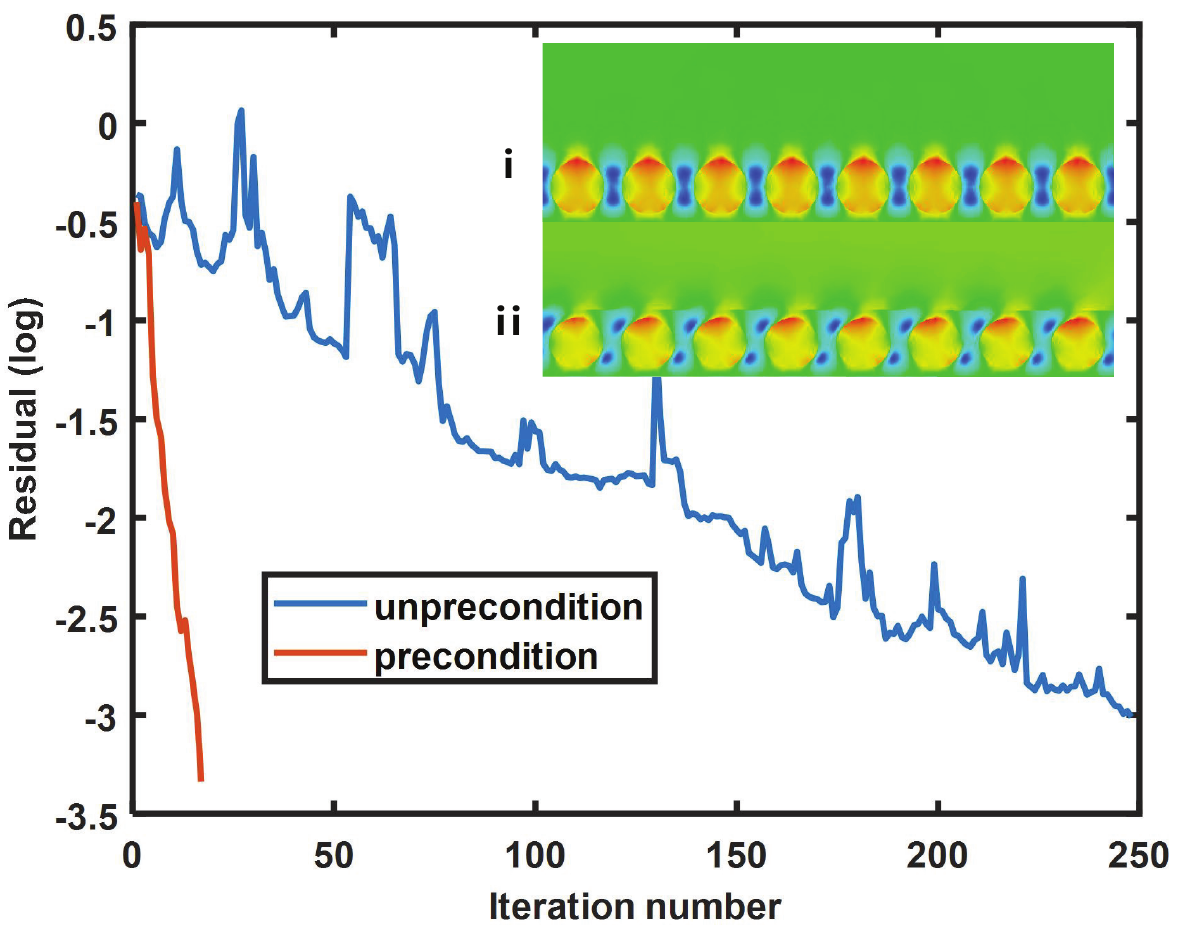}}
	\caption{Residual error versus iteration number for the OSCs.}
	\label{huang8}
\end{figure}

\begin{table*}[ht!]
	\begin{center}
		\caption{COMPUTATIONAL STATISTICS OF OSCs WITH THE RESIDUAL ERROR
				\label{table2}
		 $\epsilon=10^{-3}$}
		\begin{tabular}{| c | c | c | c | c |}
			\hline
			Method (i,ii)&Steps (\romannumeral1) & CPU time (\romannumeral1)&Steps (\romannumeral2)& CPU time (\romannumeral2)  \\			
			\hline
			original & 247  & 34 min 12 s & 276 & 35 min 35 s\\
			\hline
			x-4blocks and z-2blocks & 12  & 29 min 15 s & 16 & 34 min 25 s \\
			\hline
			x-8blocks and z-2blocks & 19  & 20 min 32 s & 16 & 18 min 59 s\\
			\hline
			x-8blocks & 12  & 34 min 35 s & 11 & 39 min 56 s\\
			\hline
		\end{tabular}
	\end{center}
\end{table*}
The organic solar cells (OSCs) are promising for future green energy applications. Considering the low carrier mobility and the short exciton diffusion length, metallic nanoparticles are embedded into OSCs for enhancing the optical absorption on the basis of local plasmon resonance\cite{32he2018osc}. As displayed in Fig. \ref{huang7}, eight identical gold nanoparticles ($\epsilon_{r}$ = $-$8.45 $-$ $j$1.41) are incorporated into a silicon spacer, which is illuminated by an $x$-polarized wave along the $z$-axis at 500 THz. The diameter of the nanoparticle is $D=$ 60 nm,  and the particle-particle spacing is $L=$ 20 nm. The thicknesses of the spacer (silicon) and the organic active layer (P3HT:PCBM) are $H1$ = 80 nm and $H2$ = 120 nm, respectively. Different from the dielectric metasurface, strong near-field evanescent wave coupling exists between the nanoparticles, which significantly slows down the iteration process. Here, we study the performances of the preconditioner under the two cases of vertical incidence and oblique incidence with different  ways of division. And the results are shown in Table \ref{table2}. In order to realize parallelization, the whole structure is divided into different longitudinal sections based on the regularity and topology. Considering the high dielectric contrast between the organic layer and spacer, division along the interface of media is also studied. The inner residual error for the stopping criteria of preconditioning is set to 0.03. The residual error of iteration and the distribution of electric field are displayed in Fig. \ref{huang8}, and strong plasmon coupling between nanoparticles is clearly observed. The inset shows the near-field distribution for \romannumeral1. vertical incidence. \romannumeral2. oblique incidence. From the Table \ref{table2}, it is obvious that block preconditioner works well in the two cases. In case \romannumeral1, the near-field energy scattered from the metal nanoparticles is mainly distributed along the polarization direction ($x$-direction) of incident field based on the wave physics of local plasmon. When we try to divide more blocks along the $x$-direction and maintain the same number of blocks along the $z$-direction, according to the analysis of the trade-off in Section \uppercase\expandafter{\romannumeral2} Part B, less field information will be contained in each block leading to coarse approximation, which increases the iteration steps. However, the reduced calculation time of submatrix equations for smaller blocks exceeds the time consumed by the increased iteration steps, achieving better preprocessing results. Similarly, in the case {\romannumeral2}, only a relatively small amount of energy penetrates into the active layer, which could be explained by the $R^{-3}$ decay of electric field and the reflection by the interface\cite{33Sha2011Near}. The local fields near the nanoparticles are fully contained in the corresponding blocks, that is the reason why the iteration steps are not reduced when the number of blocks along the $x$-direction are different. But the calculation time is reduced a lot due to the simplicity of solving the equation for smaller blocks. Therefore, we conclude that the small blocks capture field information well and therefore make the preconditioning more effective. When comparing different divisions along the vertical direction, the results show that if there is no division along the $z$-direction, the whole solution time increases a lot, which indicates the solution of submatrix equation is time consuming for the block containing high-contrast dielectric interface.
\begin{table}
	\centering
	\caption{COMPARISON OF CPU TIME BETWEEN SERIAL AND PARALLEL IMPLEMENTATIONS}
	\label{table3}
	\begin{tabular}{|c|c|c|c|} 
		\hline
		\multirow{2}{*}{Example} & \multicolumn{2}{c|}{CPU TIME } & \multirow{2}{*}{Cores}  \\ 
		\cline{2-3}
		& Serial & Parallel             &                         \\ 
		\hline
		submatrix equations in \eqref{eq10}        &     69 min   &             5 min        &              30          \\ 
		\hline
		one MVM by FFT                      &    30 s    &             18 s        &              30        \\
		\hline
	\end{tabular}
\end{table}\\
\subsection{Multithread parallel computation}   
In order to minimize the computational time as much as possible, we use OpenMP as an easy-to-use and simple programming environment for the multithread computations to accelerate the FFT. More importantly, regarding the solution to divided submatrix equations of the preconditioner in the iteration (See \eqref{eq10}), the OpenMP instruction is also employed to perform parallel computing. The acceleration of them by the OpenMP is independent of  the scheme of precondition. In order to clarify the effects of the multithread operations on the acceleration of them, the calculation time with and without the parallelization are compared in Table \ref{table3}. The results show that the multithread operation has a significant impact on the acceleration of the divided submatrx equations of the preconditioner, which is superior to the acceleration effect on the FFT. Thus, the acceleration by the OpenMP is mainly attributed to the employment of solving independent submatrix equations of the preconditioner.

\section{Conclusion}
In this work, the VIE-FFT algorithm combined with a parallel block preconditioned BiCGSTAB method is proposed for modeling the electromagnetic response from nanostructures with two representative examples of focusing metasurfaces and plasmonic solar cells. The preconditioning method employs the block decomposition technique compatible with the geometric features of nanostructures to speed up the iteration, which is suitable for OpenMP parallel implementation on distributed memory architectures. Numerical examples show that wave interaction within nanostructures influences the numerical performances of the preconditioner. The block regions need to be set to capture both far-field interference and near-field coupling effects. Consequently, the iterative steps are significantly reduced by using the developed preconditioner. In future works, expanded blocks containing more field information for nano-electromagnetic problems are worth being investigated.
\appendices

\end{document}